# Pairwise-based Multi-Attribute Decision Making Approach for Wireless Network

Ning Li, *Member, IEEE*, Jianen Yan, Zhaoxin Zhang, Alex X, Liu, *Fellow, IEEE*, Xin Yuan

*Abstract*—In the wireless network applications, such as routing decision, network selection, etc., the Multi-Attribute Decision Making (MADM) has been widely used. The MADM approach can address the multi-objective decision making issues effectively. However, when the parameters vary greatly, the traditional MADM algorithm is not effective anymore. To solve this problem, in this paper, we propose the pairwise-based MADM algorithm. In the PMADM, only two nodes' utilities are calculated and compared each time. The PMADM algorithm is much more accurate than the traditional MADM algorithm. Moreover, we also prove that our PMADM algorithm is sensitive to the parameters which vary seriously and in-sensitive to the parameters which change slightly. This property is better than that of the traditional MADM algorithm. Additionally, the PMADM algorithm is more stable than the traditional MADM algorithm. For reducing the computational complexity of the PMADM algorithm, we propose the low-complexity PMADM algorithm. For analyzing the computational complexity of the *l*PMADM algorithm, we propose the tree-based decomposing algorithm in this paper. The *l*PMADM algorithm has the same properties and performances as that of the PMADM algorithm; however, it is simpler than the PMADM algorithm. The simulation results show that the PMADM and *l*PMADM algorithms are much more effective than the traditional MADM algorithm.

*Keywords*—Multi-Attribute decision making, routing, network selection, data center network, wireless network, pairwise-based.

## I. Introduction

### A. Motivation

In many network applications, more than one parameter needs to be taken into account to choose the best candidate or sort these candidates. For instance, in deterministic routing, one next hop relay node needs to be selected based on one or more parameters [1]; in opportunistic routing, the sender needs to sort all candidate forwarding nodes based on the parameters of these nodes [2]; in heterogeneous network, the user needs to choose the best forwarding network for data forwarding [9]; in data center network, the user needs to decide which server is selected to upload data [21-22]; etc. At present, there are two different kinds of algorithms that can address this kind of problems: the traditional MADM based approach and the fuzzy logic based approach. The fuzzy logic based approach is accurate. However, on one hand, its computational complexity is high; on the other hand, with the increasing of the number of parameters and the number of nodes, the computational complexity increases exponentially in the fuzzy logic based approach [3-5]. This is not appropriate for the wireless nodes, in which the compute capacity is always limited. The traditional MADM based approach is applied widely because of its simplicity and effectiveness, such as the simple additive weighting approach (SAW) [6][7], the weighting product approach (WP) [6][7], the technique for order preference by similarity to ideal solution approach (TOPSIS) [6][7], etc. However, the traditional MADM based approach is not as accurate as that of the fuzzy logic based approach. Therefore, we propose the most accurate MADM based approach in this paper to solve this kind of issue, shorted as PMADM and *l*PMADM.

### B. Problems Statement

The main problem needs to be solved in this paper is to improve the accuracy and stability of the traditional MADM based approach. Specifically, the problems that need to be solved in this paper can be summarized as follows.

First, for the traditional MADM approach, the most important thing is to decide the weight of each parameter [7]. Since for different weight calculation algorithms, the utilities of the same node will be different, i.e., the final decision will be different. For different weight calculation algorithms, the same MADM algorithm may make different decisions to the same problem. Therefore, a fair and effective weight calculation algorithm is necessary.

TABLE 1. AN EXAMPLE

| $P$   | N1    | N2    | N3    | N4    | N5    |
|-------|-------|-------|-------|-------|-------|
| $P_1$ | 0.1   | 0.2   | 0.3   | 0.4   | 0.9   |
| $P_2$ | $b_1$ | $b_2$ | $b_3$ | $b_4$ | $b_5$ |
| $P_3$ | $c_1$ | $c_2$ | $c_3$ | $c_4$ | $c_5$ |
| $P_4$ | $d_1$ | $d_2$ | $d_3$ | $d_4$ | $d_5$ |

Second, another very important and common issue in the MADM approach is described as follows. In most cases, the traditional MADM approach is effective. However, in the scenario that the same parameter varies greatly with different nodes, the traditional MADM approach is not effective anymore, which can be explained in Table 1. In Table 1, there are five nodes and four parameters in each node; the value of $P_1$ in N5 is much larger than that of the other four nodes, and the other four nodes are similar. In this scenario, the traditional MADM algorithm does not work precisely. This example only introduces when one parameter has this kind of issue; when more than one parameter has this kind of issue, the calculation becomes much more difficult and complex.

Third, similar to the second problem, because the values of parameters have a great effect on the weights, the changing of parameter value may cause the changing of the nodes' sequences. However, a good decision making algorithm should not change the relative nodes sequences whose parameters are not changed. For instance, there are five nodes: N1, N2, N3, N4, and N5; the sequence of these five nodes is N1>N2>N3>N4>N5. If one of the parameters in N5 changes and makes the utility of N5 larger than that of N3 and N4, then a good decision making algorithm should guarantee that the sequence is N1>N2>N5>N3>N4, i.e., the relative sequences of N1, N2, N3, and N4 do not change. This issue will be solved in this paper.

### C. Limitations of prior arts

There are limitations in the prior arts. For the fuzzy logic based approach, the first limitation is its high computational complexity. So, it is not suitable for the wireless nodes, in which the memory space and the compute capability is extremely limited, such as the nodes in WSNs. Second, as introduced in [3-5], with the increasing of the number of parameters and the number of nodes, the computational complexity increases exponentially. This will restrict the application of fuzzy logic based approach in wireless network; especially when the number of parameters and the number of nodes are large. In the previous works, the general number of parameters is 3 in the fuzzy logic based approach [3].

For the traditional MADM approach, first, the parameters in different nodes are close-coupled. This means that when one node changes its parameter, then all the other nodes will be affected. This is not appropriate since for a good decision making algorithm, when one node's parameter changes, only the relative sequence of this node is affected; the sequence of the other nodes should not be affected. Second, the accuracy of the traditional MADM based approach is not as good as that of the fuzzy logic based approach. Third, for the conventional approach, when one parameter in the node changes, the utilities of all the nodes need to be re-calculated, which is complicated and time wasted.

*D. Proposed approach and advantages over prior arts*

Based on the limitations and problems introduced above, in this paper, firstly, we propose a variance-based weight calculation algorithm for the MADM approach; secondly, based on this weight calculation algorithm, we propose the pairwise-based MADM algorithm (PMADM) and the low-complexity pairwise-based MADM (*l*PMADM) algorithm. In the weight calculation algorithm, according to the conclusions in our previous works, the node's parameters whose variances are larger have larger effects on the final decision than the parameters whose variances are smaller. Therefore, in PMADM, the parameter whose variance is large has a larger weight than the parameter whose variance is small. In the PMADM algorithm, we do not calculate the weights of parameters and the utilities of all nodes directly, we compute the weights and utilities of only two nodes each time, i.e., we calculate the weights of parameters and the utilities of nodes pairwise. In each calculation, the algorithm is the traditional MADM algorithm. Since for the PMADM algorithm, the times of calculation is large, for reducing the computational complexity, we propose the *l*PMADM. In *l*PMADM, if the sequence of N$i$ is $k$, then there are $k-1$ nodes whose utilities are smaller than N$i$, and $m-k$ nodes whose utilities are bigger than N$i$. Moreover, the utilities of these $k-1$ nodes (whose utilities are smaller than N$i$) are also smaller than the other $m-k$ nodes (whose utilities are larger than N$i$). Thus, the computational complexity is reduced greatly.

The proposed PMADM and *l*PMADM algorithms have advantages over the prior arts. First, the PMADM/*l*PMADM algorithm is much simpler than the fuzzy logic based algorithm; additionally, the proposed PMADM/*l*PMADM algorithm is much more effective than the traditional MADM algorithm. Second, in the PMADM/*l*PMADM algorithm, with the increasing of the number of parameters and the number of nodes, the increasing of the computational complexity is lower than that of the fuzzy logic based algorithm. Third, in the proposed PMADM/*l*PMADM algorithm, the parameters in different nodes are loosely coupled. When one node changes its parameter, only the relative sequence of this node is changed; the relative sequences of the other nodes are constant. Finally, in the proposed PMADM/*l*PMADM algorithm, when one of the parameters in the nodes is changed, only the utility of this node needs to be re-calculated, the utilities of the other nodes do not need to be calculated. This is simpler than the traditional MADM algorithm.

## II. RELATED WORKS

The MADM algorithm has been used in wireless networks for different purposes in the past decades. In [8], the authors introduce the MADM algorithm into the routing selection in the opportunistic routing of VAENTs. In this algorithm, the Signal to Interference plus Noise Ratio (SINR) and the packet queue length are taken into account by the MADM algorithm to decide the node utility. In [9], the authors use the MADM algorithm to select the best interface dynamically. In this algorithm, the interface characteristics and user preferences are taken into account. In [10], for the next hop node selection, the authors take the congestion and routing paradigm into account and apply the MADM to choose the best next hop relay node. In [11], the authors apply the MADM algorithm into the source-based path selection strategy for the multi-hop wireless networks. By this approach, the overall performance of such networks by exploiting paths traditionally disregarded by the routing protocols is improved. In [12], for solving the routing issues in LEO-satellite DTN network, the authors take the link bandwidth, the link establishment delay, the free storage space at nodes, network BER, and data forwarding rate into account to realize the DTN routing. In [13], the authors propose a new adaptive and dynamic multi-constraint routing approach. In this approach, the network topology, the power consumption and the residual energy of nodes are considered to build the routing tree. Moreover, a dynamic selection mechanism based on MADM approach is implemented to build and update the routing tree. In the next generation network, the network selection will be based on multi standards. So the authors in [14] introduce the MADM approach into the network selection based on multiple network parameters to improve the performances of selection. In [15], the authors formulate the interest packet forwarding as a multiple attribute decision making problem; the authors proposed the forwarding strategy model based on this fact. In [16], for improving the performances of the proposed algorithm in [15], the Ant Colony Optimization is introduced into the algorithm in [15]. Based on these innovations, the proposed algorithm can improve the packet forwarding performance in NDN. For addressing the routing issue in the VANETs in which there are many outstanding solutions available, in [17], the authors regard a vehicle as a decision maker and introduce the MADM approach into the routing decision to enhance the routing performances.

## III. THE PMADM ALGORITHM

In this section, we will introduce the proposed PMADM algorithm in detail. Before introducing the PMADM algorithm, we first propose the variance based weight calculation algorithm.

*A. Variance based weight calculation algorithm*

For the MADM algorithms, according to the conclusions in our previous works, we find that the node's parameters whose variances are large have great effects on the final decision. We take the next hop relay node selection in routing algorithm as an example. For instance, suppose there are three nodes and two parameters in each node: $P_1$ and $P_2$. The values of $P_1$ in these three nodes are 101, 102, and 103, respectively. The values of $P_2$ of these three nodes are 0.1, 0.5, and 0.9, respectively. So, intuitively, for $P_1$, which node is chosen has a small effect on the routing performance; however, for $P_2$, which node is selected has a great effect on the routing performance. To address this problem, we introduce to use the variances of parameters to calculate the parameters' weights in this paper.

For each node, there are $m$ candidate nodes for the next hop selection and $n$ parameters are taken into account for each node. Therefore, the set of parameters of node $i$ is $\boldsymbol{P}_i = \{P_i^1, P_i^2, P_i^3, \ldots, P_i^n\}$ and the set of nodes is $\boldsymbol{M} = \{1, 2, 3, \ldots, m\}$. Then the parameter matrix of all the node is $\boldsymbol{NM} = \begin{bmatrix} P_1^1, P_1^2, P_1^3, \ldots, P_1^n \\ P_2^1, P_2^2, P_2^3, \ldots, P_2^n \\ P_3^1, P_3^2, P_3^3, \ldots, P_3^n \\ \vdots \\ P_m^1, P_m^2, P_m^3, \ldots, P_m^n \end{bmatrix}$, where $P_i^j$ means the $j$th parameter in N$i$.

Based on the principle of MADM [7], the first step is the normalization of all the parameters in $\boldsymbol{NM}$. There are many parameter normalization algorithms, such as the SAW [6][7], the WP [6][7], etc. The purpose of parameter normalization is to make all the parameters in the same order of magnitude [6][7]. After parameter normalization, the parameter matrix can be expressed as:

$$\boldsymbol{NM}^* = \begin{bmatrix} P_1^{1*}, P_1^{2*}, P_1^{3*}, \ldots, P_1^{n*} \\ P_2^{1*}, P_2^{2*}, P_2^{3*}, \ldots, P_2^{n*} \\ P_3^{1*}, P_3^{2*}, P_3^{3*}, \ldots, P_3^{n*} \\ \vdots \\ P_m^{1*}, P_m^{2*}, P_m^{3*}, \ldots, P_m^{n*} \end{bmatrix} \quad (1)$$

Thus, the variance of each parameter can be calculated as:

$$v_i = \frac{\sum_{j=1}^{m}\left(P_i^j - \bar{P}_i\right)^2}{m} \quad (2)$$

where $1 < i < n$ is the index of different parameters, $1 < j < m$ is the index of different nodes. When we get the variance of each parameter, these variances cannot be used as the weight directly, because in MADM algorithm, the sum of weights should equal to 1 [6][7]. Therefore, we calculate the weight of each parameter based on the variance as:

$$\omega_i = \frac{v_i}{\sum_{i=1}^{n} v_i} \quad (3)$$

*B. Pairwise-based MADM algorithm*

In this paper, to address the problems introduced in Section I, we proposed the pairwise-based MADM algorithm. The PMADM algorithm is simple and effective. The main novelty of the PMADM algorithm is that we calculate and compare the utilities of nodes pairwise, i.e., we calculate and compare the utilities of only two nodes each time.

Suppose there are $m$ nodes, which are N1, N2, ...., and N$m$, respectively. Firstly, we compute the utilities between N1 and N2. The calculation is based on the traditional MADM algorithm and the weights of the parameters in N1 and N2 are decided according to the algorithm introduced in Section III.A. This process will be repeated until the utilities between any two nodes are calculated. Note that in PMADM, when calculating the utilities between two nodes, only these two nodes are involved; the other nodes are not involved in the utility computation of these two nodes. This is different from the traditional MADM algorithm. Consequently, in PMADM, we calculate the variances of only two nodes' parameters each time rather than the variances of all nodes' parameters. Thus, for the same parameter, there will be more than one variances in PMADM algorithm. So, we define the variance matrix for each parameter as follows.

**Definition 1.** The variance matrix of parameter $i$ is defined as the set of the variances of parameter $i$ between different node pairs, denoted as $v_i^*$; the $v_i^*$ can be expressed as:

$$v_i^* = \begin{bmatrix} v_i^{11} & v_i^{12} & \cdots & v_i^{1m} \\ v_i^{21} & v_i^{22} & \cdots & v_i^{2m} \\ \vdots & \vdots & \vdots & \vdots \\ v_i^{m1} & v_i^{m2} & \cdots & v_i^{mm} \end{bmatrix} \quad (4)$$

In (4), $m$ is the number of nodes to be selected; $v_i^{12}$ means the variance of parameter $i$ between N1 and N2. Moreover, in (4), $v_i^{11} = 0$ and $v_i^{21} = v_i^{12}$. Based on the Definition 1, the variances of all the other $n$ parameters can be calculated. For instance, $v_1^*$ is shown as follows.

$$v_1^* = \begin{bmatrix} v_1^{11} & v_1^{12} & \cdots & v_1^{1m} \\ v_1^{21} & v_1^{22} & \cdots & v_1^{2m} \\ \vdots & \vdots & \vdots & \vdots \\ v_1^{m1} & v_1^{m2} & \cdots & v_1^{mm} \end{bmatrix} \quad (5)$$

Based on the variance matrix and (3), the weights of each parameter between any two nodes can be calculated. For the PMADM algorithm, we compute the utilities of nodes pairwise. Then the utilities of N$a$ and N$b$ can be calculated as:

$$U_{ab}^a = \sum_{i=1}^{n} \omega_i^{ab} \cdot p_i^a \quad (6)$$

$$U_{ab}^b = \sum_{i=1}^{n} \omega_i^{ab} \cdot p_i^b \quad (7)$$

In (6) and (7), $\omega_i^{ab}$ is the weight of parameter $i$ between N$a$ and N$b$; $P_i^a$ is the value of parameter $i$ in N$a$; $U_{ab}^a$ is the utility of N$a$ when compared with N$b$. From (6) and (7), we can find that in PMADM, the utility is meaningful only when we specify which two nodes that are compared. This is because even for the same node, when compared with different nodes, its utility will be different. For instance, for N1 and N2, the utility of these two nodes can be calculated as $U_{12}^1 = \sum_{i=1}^{n} \omega_i^{12} \cdot P_i^1$ and $U_{12}^2 = \sum_{i=1}^{n} \omega_i^{12} \cdot P_i^2$. For N1 and N3, the utilities can be calculated as $U_{13}^1 = \sum_{i=1}^{n} \omega_i^{13} \cdot P_i^1$ and $U_{13}^3 = \sum_{i=1}^{n} \omega_i^{13} \cdot P_i^3$. The values of utilities $U_{12}^1$ and $U_{13}^1$ are different, even they are all the utilities of N1. Therefore, if the PMADM is effective, the following property should be satisfied: when $U_{12}^1 > U_{12}^2$ and $U_{23}^2 > U_{23}^3$, then the algorithm should guarantee that $U_{12}^1 > U_{23}^3$, i.e., the result is not a loop. Thus, we give the corollary as follows.

**Corollary 1.** In the PMADM algorithm, the result is not a loop; i.e., if $U_{ij}^i > U_{ij}^j$ and $U_{jk}^j > U_{jk}^k$, then $U_{ik}^i > U_{ik}^k$ holds.

*Proof.* To simplify, we assume that there are four parameters for each node, which are $P_A$, $P_B$, $P_C$, and $P_D$, respectively. Moreover, $U_{12}^1 > U_{12}^2$ and $U_{23}^2 > U_{23}^3$. For the PMADM algorithm, the utilities $U_{12}^1$ and $U_{12}^2$ can be calculated based on (6) and (7), respectively, which are:

$$U_{12}^1 = \omega_A^{12} P_A^1 + \omega_B^{12} P_B^1 + \omega_C^{12} P_C^1 + \omega_D^{12} P_D^1 \quad (8)$$

$$U_{12}^2 = \omega_A^{12} P_A^2 + \omega_B^{12} P_B^2 + \omega_C^{12} P_C^2 + \omega_D^{12} P_D^2 \tag{9}$$

Thus, $U_{12}^{12}$ can be computed as:

$$\begin{aligned} U_{12}^{12} &= U_{12}^1 - U_{12}^2 \\ &= \omega_A^{12} P_A^{12} + \omega_B^{12} P_B^{12} + \omega_C^{12} P_C^{12} + \omega_D^{12} P_D^{12} \end{aligned} \tag{10}$$

In the PMADM, the weight calculation algorithm is the same as that shown in Section III.A. Moreover, the variance of $P_A^1$ and $P_A^2$ can be calculated as:

$$v_A^{12} = \tfrac{1}{4}(P_A^{12})^2 \tag{11}$$

Similarly, $v_B^{12} = \tfrac{1}{4}(P_B^{12})^2$, $v_C^{12} = \tfrac{1}{4}(P_C^{12})^2$, and $v_D^{12} = \tfrac{1}{4}(P_D^{12})^2$. According to the weight calculation algorithm, since $\omega_A^{12} + \omega_B^{12} + \omega_B^{12} + \omega_B^{12} = 1$, we can calculate the weight of each parameter as follows.

$$\omega_A^{12} = \frac{v_A^{12}}{v_A^{12}+v_B^{12}+v_C^{12}+v_D^{12}} = \frac{(P_A^{12})^2}{(P_A^{12})^2+(P_B^{12})^2+(P_C^{12})^2+(P_D^{12})^2} \tag{12}$$

$$\omega_B^{12} = \frac{v_B^{12}}{v_A^{12}+v_B^{12}+v_C^{12}+v_D^{12}} = \frac{(P_B^{12})^2}{(P_A^{12})^2+(P_B^{12})^2+(P_C^{12})^2+(P_D^{12})^2} \tag{13}$$

$$\omega_C^{12} = \frac{v_C^{12}}{v_A^{12}+v_B^{12}+v_C^{12}+v_D^{12}} = \frac{(P_C^{12})^2}{(P_A^{12})^2+(P_B^{12})^2+(P_C^{12})^2+(P_D^{12})^2} \tag{14}$$

$$\omega_D^{12} = \frac{v_D^{12}}{v_A^{12}+v_B^{12}+v_C^{12}+v_D^{12}} = \frac{(P_D^{12})^2}{(P_A^{12})^2+(P_B^{12})^2+(P_C^{12})^2+(P_D^{12})^2} \tag{15}$$

Thus, the (10) can be rewritten as:

$$\begin{aligned} U_{12}^{12} &= U_{12}^1 - U_{12}^2 \\ &= \frac{1}{(P_A^{12})^2+(P_B^{12})^2+(P_C^{12})^2+(P_D^{12})^2} \\ &\quad \cdot [(P_A^{12})^3 + (P_B^{12})^3 + (P_C^{12})^3 + (P_D^{12})^3] \end{aligned} \tag{16}$$

Since $U_{12}^{12} > 0$ and $\frac{1}{(P_A^{12})^2+(P_B^{12})^2+(P_C^{12})^2+(P_D^{12})^2} > 0$, $[(P_A^{12})^3 + (P_B^{12})^3 + (P_C^{12})^3 + (P_D^{12})^3] > 0$ holds. Similar to the calculation in (16), the $U_{23}^{23}$ can be calculated as:

$$\begin{aligned} U_{23}^{23} &= U_{23}^2 - U_{23}^3 \\ &= \frac{1}{(P_A^{23})^2+(P_B^{23})^2+(P_C^{23})^2+(P_D^{23})^2} \\ &\quad \cdot [(P_A^{23})^3 + (P_B^{23})^3 + (P_C^{23})^3 + (P_D^{23})^3] \end{aligned} \tag{17}$$

Since $U_{23}^{23} > 0$ and $\frac{1}{(P_A^{23})^2+(P_B^{23})^2+(P_C^{23})^2+(P_D^{23})^2} > 0$, $[(P_A^{23})^3 + (P_B^{23})^3 + (P_C^{23})^3 + (P_D^{23})^3] > 0$ holds. The $U_{13}^{13}$ can be derived as:

$$\begin{aligned} U_{13}^{13} &= U_{13}^1 - U_{13}^3 \\ &= \frac{1}{(P_A^{13})^2+(P_B^{13})^2+(P_C^{13})^2+(P_D^{13})^2} \\ &\quad \cdot [(P_A^{13})^3 + (P_B^{13})^3 + (P_C^{13})^3 + (P_D^{13})^3] \end{aligned} \tag{18}$$

So, we can conclude that:

$$\begin{aligned} &U_{12}^{12} + U_{23}^{23} \\ &> K \cdot \{[(P_A^{12})^3 + (P_A^{23})^3] + [(P_B^{12})^3 + (P_B^{23})^3] \\ &\quad + [(P_C^{12})^3 + (P_C^{23})^3] + [(P_D^{12})^3 + (P_D^{23})^3]\} \end{aligned} \tag{19}$$

where $K = \min\left(\frac{1}{(P_A^{12})^2+(P_B^{12})^2+(P_C^{12})^2+(P_D^{12})^2}, \frac{1}{(P_A^{23})^2+(P_B^{23})^2+(P_C^{23})^2+(P_D^{23})^2}\right)$.

What's more, since $P_A^{12} = P_A^1 - P_A^2$, $P_A^{23} = P_A^2 - P_A^3$, so $P_A^{13} = P_A^{12} + P_A^{23} = P_A^1 - P_A^3$. Moreover, since $U_{12}^{12} > 0$ and $U_{23}^{23} > 0$. according to the conclusion that $x^3 + x^3 < (x+y)^3$, we have:

$$U_{12}^{12} + U_{23}^{23} < M \cdot [(P_A^{13})^3 + (P_B^{13})^3 + (P_C^{13})^3 + (P_D^{13})^3] \tag{20}$$

where $M = \max\left(\frac{1}{(P_A^{12})^2+(P_B^{12})^2+(P_C^{12})^2+(P_D^{12})^2}, \frac{1}{(P_A^{23})^2+(P_B^{23})^2+(P_C^{23})^2+(P_D^{23})^2}\right)$.

Since $M > 0$, so $[(P_A^{13})^3 + (P_B^{13})^3 + (P_C^{13})^3 + (P_D^{13})^3] > 0$; moreover, $\frac{1}{(P_A^{13})^2+(P_B^{13})^2+(P_C^{13})^2+(P_D^{13})^2} > 0$, so $U_{13}^{13} = U_{13}^1 - U_{13}^3 > 0$. Thus, Corollary 1 holds. ∎

However, the PMADM algorithm has higher computational complexity than that of the traditional MADM algorithm. The complexity of MADM and PMADM is shown as follows.

**Corollary 2.** The utility calculation times in PMADM algorithm are $\frac{m(m-1)}{2}$, where $m$ is the number of nodes.

*Proof.* For the PMADM algorithm, the utility calculation is between any two different nodes. So, for the first node, the utility calculation times will be $(m-1)$; for the second node, this number is $(m-2)$, and so on. The utility calculation times in the second final node is 1 and is 0 in the final node. This is an arithmetic progression. Therefore, the calculation times can be calculated as:

$$N = m \cdot 0 + \tfrac{m(m-1)}{2} = \tfrac{m(m-1)}{2} \tag{20}$$

Thus, the Corollary 2 holds. ∎

Based on the Corollary 2, we give the computational complexity of both the MADM algorithm and the PMADM algorithm.

**Corollary 3.** The computational complexity of the traditional MADM algorithm is $O(nm)$ and the computational complexity of the PMADM algorithm is $O[nm^2]$.

*Proof.* There are $m$ nodes and $n$ parameters in each node. For the MADM algorithm, first, it needs to calculate the weight of each parameter, so the complexity of this step is $O(n)$. After that, the weight needs to multiply by the parameter and calculate the sum of the results of multiplication, so the complexity of this step is also $O(n)$. Finally, for each node, this process will be repeated. Since there are $m$ nodes, the computational complexity of the MADM algorithm is $O(nm)$.

For the PMADM algorithm, first, it needs to calculate the weight of each parameter, so the complexity of this step is $O(n)$; after that, the weight needs to multiply by the parameter and calculate the sum of the result of multiplication, so the complexity of this step is also $O(n)$. However, different from the PMADM algorithm, the utility of each node will be calculated between only two nodes each time, and this process will be repeated $\frac{m(m-1)}{2}$ times. So, the computational complexity of this step is $O(m^2)$. Thus, the final computational complexity of PMADM algorithm is $O[nm^2]$. ∎

From the Corollary 3, we can conclude that: first, the computational complexity of PMADM is higher than that of MADM; second, with the increasing of *n* or *m*, the computational complexity of PMADM increases much more quickly than MADM; third, in MADM, when *n* or *m* increases, the computational complexity increase equally; however, in PMADM, the computational complexity increases much more quickly with the increasing of *m* than *n*.

## IV. THE LOW-COMPLEXITY PMADM ALGORITHM

From Corollary 3, we can conclude that the PMADM has limited capability on handling the scenario that there are many candidate nodes. Therefore, for improving the performances and reducing the computation complexity of PMADM algorithm, we propose the low-complexity PMADM algorithm, shorted as *l*PMADM.

### A. Low-complexity PAMDM

Assuming that based on Corollary 1, the sequence of N1 is *k*. For N1, its utilities relative to other nodes are: $U_{12}^1, U_{13}^1, U_{14}^1, \ldots, U_{1m}^1$; the utilities of the other nodes relative to N1 are: $U_{12}^2, U_{13}^3, U_{14}^4, \ldots, U_{1m}^m$. If the sequence of N1 is *k*, it means that there are $k-1$ nodes that can satisfy $U_{1i}^1 > U_{1i}^i$, where $1 < i < m$; moreover, there are $m-k$ nodes that satisfy that $U_{1j}^1 < U_{1j}^j$, where $1 < j < m$. Assume that the utilities of N2, N3, …., N*k* are smaller than that of N1; the utilities of N(*k*+1), N(*k*+2), …, N*m* are larger than that of N1. So, based on the conclusion in Corollary 1, the utilities of N2, N3, …., N*k* are smaller that of N(*k*+1), N(*k*+2), …, N*m*. This illustrates that we do not need to compare the utilities of N2, N3, …., N*k* with N(*k*+1), N(*k*+2), …, N*m* anymore. What we need to do is to compare the utilities between N2, N3, …., and N*k* as the same process of N1. The same as N(*k*+1), N(*k*+2), …, and N*m*. So, we have the corollary as follows.

**Corollary 4.** Based on the PMADM algorithm, if the sequence of N*i* is *k*, then there are $k-1$ nodes whose utilities are smaller than N*i*; moreover, there are $m-k$ nodes whose utilities are larger than N*i*. Besides, the utilities of these $k-1$ nodes whose utilities are smaller than N*i* are also smaller than the other $m-k$ nodes whose utilities are larger than N*i*.

*Proof.* According to the conclusion in the Corollary 1, the proposed PMADM algorithm is loop-free, so if the sequence of N*i* is *k*, then there must exit $k-1$ nodes whose utilities are smaller than N*i*, and there are $m-k$ nodes whose utilities are larger than N*i*. Since the PMADM algorithm is loop-free, so the utilities of these $k-1$ nodes must be smaller than the other $m-k$ nodes. ∎

The Corollary 4 shows that it is possible to reduce the computational complexity of the PMADM algorithm. For instance, firstly, choosing the first starting node randomly; secondly, calculating the utility and sequence of this node, denoted as *k*; thirdly, repeating Step 1 to Step 2; calculating the utilities of the nodes whose utilities are smaller than *k* and the utilities of the nodes whose utilities are larger than *k*, respectively; finally, repeating Step 1 to Step 3, until all the nodes' utilities are calculated. Based on these steps, the computational complexity of PMADM can be reduced to some extent. However, if we do not choose the starting node randomly, then the computation complexity of PMADM can be reduced further. Before that, we introduce one important property of PAMDM in Corollary 5.

**Corollary 5.** In the tree-based decomposing algorithm, if in each step, the sequence of the starting node is the middle one, then the number of layers and the comparison times are the smallest. i.e., if *m* is odd, then $k = \frac{m+1}{2}$; if *m* is even, then $k = \frac{m}{2}$ or $k = \frac{m+2}{2}$.

*Proof.* In Corollary 5, the comparison times is the number of comparisons between any two nodes. Since the proposed PMADM algorithm is similar to the QuickSort algorithm and the computational complexity of the QuickSort algorithm is the smallest when the starting node is the middle one, the Corollary 5 holds. ∎

However, even our proposed PMADM algorithm is similar to the QuickSort algorithm, they are very different. Inspired by Corollary 5, if we can choose the starting node satisfies the requirements in Corollary 5, then the computational complexity of PMADM algorithm can be reduced further. We call this is the low-complexity PMADM, shorted as *l*PMADM. However, in practice, the value of *k* does not always satisfy the requirements of Corollary 5. Consequently, we propose two pre-sequencing approaches to increase the probability that the chosen starting node satisfies the requirements of Corollary 5.

The *first approach* is to use the traditional MADM algorithm to pre-calculating the sequence of each node. Then, we select the node whose sequence is the middle as the starting node in each step.

The *second approach* is shown in Table 2. For each parameter, the sequences of the parameters in different nodes are different, which are shown in Table 2. For instance, the sequence of P1 in N1 is $o_1^1$, the sequence of P2 in N1 is $o_2^1$, the sequence of P3 in N1 is $o_3^1$, and so on. The average parameter sequence of node *i* can be calculated as:

$$O_i = \frac{1}{n}\sum_{j=1}^n o_j^i \qquad (21)$$

In (21), *i* is the index of node and *j* is the index of parameter; moreover, $0 < i < m$ and $0 < j < n$. Then, we select the node whose average sequence is the middle as the starting node.

The first approach is more accurate than the second approach, since in the second approach, there may exist more than one node which has the same average sequence. However, the computational complexity of the first approach is more serious than the second approach.

TABLE 2. AN EXAMPLE FOR THE SECOND APPROACH

| P | N1 | N2 | N3 | N4 | N5 |
|---|---|---|---|---|---|
| P1 | $P_1^1$ | $P_1^2$ | $P_1^3$ | $P_1^4$ | $P_1^5$ |
| order | $o_1^1$ | $o_1^2$ | $o_1^3$ | $o_1^4$ | $o_1^5$ |
| P2 | $P_2^1$ | $P_2^2$ | $P_2^3$ | $P_2^4$ | $P_2^5$ |
| order | $o_2^1$ | $o_2^2$ | $o_2^3$ | $o_2^4$ | $o_2^5$ |
| P3 | $P_3^1$ | $P_3^2$ | $P_3^3$ | $P_3^4$ | $P_3^5$ |
| order | $o_3^1$ | $o_3^2$ | $o_3^3$ | $o_3^4$ | $o_3^5$ |
| P4 order | $P_4^1$ | $P_4^2$ | $P_4^3$ | $P_4^4$ | $P_4^5$ |
| | $o_4^1$ | $o_4^2$ | $o_4^3$ | $o_4^4$ | $o_4^5$ |
| Average order | $O_1$ | $O_2$ | $O_3$ | $O_4$ | $O_5$ |

The low-complexity PMADM can be expressed as follows.

Step 1. Choosing the starting nodes based on the proposed two approaches;

Step 2. Calculating the utility and sequence of this node, denoted as $k$;

Step 3. Repeating the Step 1 to Step 2; calculating the utilities of the nodes whose utilities are smaller than $k$ and the utilities of the nodes whose utilities are larger than $k$, respectively;

Step 4. Repeating the Step 1 to Step 3, until all the utilities of nodes are calculated.

*B. Computational Complexity of low-complexity PMADM*

For investigating the computational complexity of low-complexity PMADM, we proposed the decomposing tree, which is shown in Fig.1.

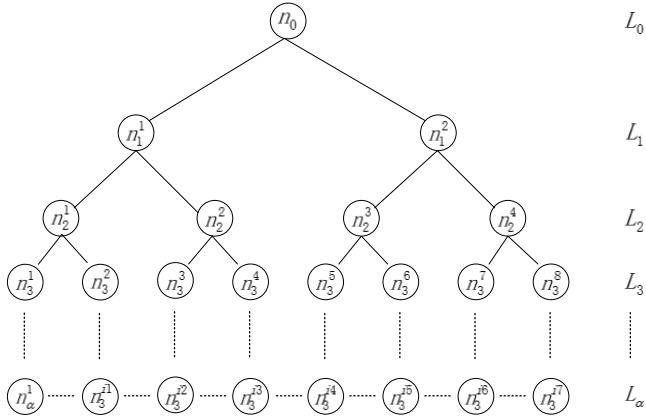

Fig.1 The decomposing tree

In the decomposing tree, the child-node is the node which is not in layer 0; the parent-node is the node which is not in the final layer. In the decomposing tree, the final layer does not mean the last layer of the decomposing tree, it is the layers after which there are no nodes anymore. For instance, in Fig.2, the layers in the red ellipse in both layer 3 and layer 4 are the final layers; only the layer in the blue rectangle in layer 4 is the last layer of the decomposing tree. So, for the same nodes, they can be parent-node and child-node at the same time. For instance, in Fig.1, the node $n_1^1$ is the child-node of $n_0$, at the same time, it is also the parent-node of $n_2^1$ and $n_2^2$. However, in the decomposing tree, the node in layer 0 is the parent-node only, while the nodes in the final layer are the child-node only. From Fig.2, we can conclude that when the value of the node is 0 or 1, it must be the child-node. This is because the nodes' values in the final layer are all 0 or 1. However, when the node values are all 0 or 1 in this layer, it could be the final layer or last layer. In the decomposing tree, each child-node only has one parent-node, and each parent node only has two child-nodes. Moreover, the value of child-node may equal to 0. For distinguishing the different kinds of nodes in the decomposing tree, we define the degree of each node in Definition 2.

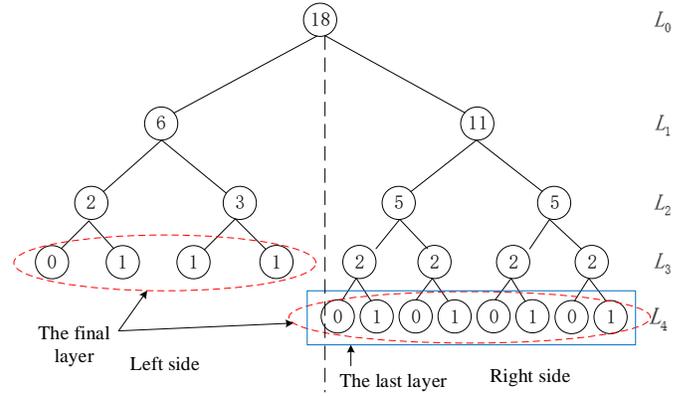

Fig.2 An example of the decomposing tree

**Definition 2.** The node degree in the decomposing tree is defined as the number of neighbor nodes of this node, including both the parent-node and child-node.

Based on Definition 2, the node degree in layer 0 is 2; the node degree in the final layer is 1; the node degree in other layers is 3. So, we use the $n$-degree node in the decomposing tree to represent different kinds of nodes, where $n = 1,2,3$. In the decomposing tree, there is only one 2-degree node and this node is always the parent-node; the 1-degree node is always the child-node and their values are 0 or 1. The role of the 3-degree nodes are depended, they can be the child-node of the higher layer nodes and they also can be the parent-node of the lower layer nodes. Moreover, the values of the 3-degree nodes are all larger than 1. The example shown in Fig.2 is useful to understand the meanings introduced above. In the decomposing tree, the parent-node and child-node have the relationship as:

$$n_i^j = n_{i+1}^{2j-1} + n_{i+1}^{2j} - 1 \qquad (22)$$

where $n_i^j$ is the value of $j$th node in layer $i$; $n_{i+1}^{2j-1}$ is the value of $(2^j - 1)$th node in layer $i + 1$; moreover, $n_i^j$ is the parent-node of $n_{i+1}^{2j-1}$ and $n_{i+1}^{2j}$. For nodes in the decomposing tree, its value should not be smaller than 0. During the decomposing, when the value of node equal to 0 or 1, the decomposing of this node will finish. For the decomposing tree, it has properties as follows.

*Number of layers and divisions.* In the following, we will investigate the number of layers and the number of divisions of the decomposing tree. The number of divisions is defined as the number of choosing the starting nodes in the *l*PMADM. For instance, there are nine nodes as N1, N2, N3, N4, N5, N6, N7, N8, and N9, respectively; suppose the nodes' utilities are consistent with their index. If in the first choice, N5 is chosen, then the whole node set will be divided into two parts: (N1, N2, N3, N4) and (N6, N7, N8, N9). For the second step, we will choose the starting node of these two sets respectively. Assuming that the N2 is chosen in the second choice and N7 is chosen in the third choice. So, the number of divisions of (N1, N2, N3, N4, N5, N6, N7, N8, N9) is 3.

**Corollary 6.** For the decomposing tree, if in each step, the starting node satisfies the requirements of Corollary 5, then there are $\lfloor \log_2 m \rfloor$ layers and $2^{\lfloor \log_2 m \rfloor - 1} - 1$ divisions in each decomposing tree.

*Proof.* This can be proved based on the QuickSort algorithm. ∎

Based on Corollary 5 and Corollary 6, we have Corollary 7 as follows.

**Corollary 7.** There are at least $\lfloor \log_2 m \rfloor$ layers and $2^{\lfloor \log_2 m \rfloor - 1} - 1$ divisions in the decomposing tree; there are at most $(m-1)$ layers and divisions in the decomposing tree.

*Proof.* For the optimal decomposing, the numbers of layers and divisions are the smallest, which are $\lfloor \log_2 m \rfloor$ and $2^{\lfloor \log_2 m \rfloor - 1} - 1$, respectively. If in each time, the starting node is the smallest one or the largest one, then only one node can be divided each time, so the higher-bounds of the numbers of layers and divisions are both $(m-1)$. ∎

*Comparison times.* Since in $l$PMADM, we need to compare the node utilities between any two nodes, the comparison times may larger than the traditional MADM approach. In the following, we will investigate the comparison times of $l$PMADM.

**Corollary 8.** Based on the tree-based decomposing algorithm, the comparison times in $l$PMADM can be calculated as:

$$c = \sum_{i=1}^{l} c_{i+1} = \sum_{i=1}^{l} \sum_{j=1}^{2^i} n_i^j \quad (23)$$

where $l$ is the number of layers.

*Proof.* In each layer of the decomposing tree, the sum of the nodes' values in this layer is the comparison times of its upper layer. For instance, in Fig.1, the comparison times of $ith$ layer is the sum of nodes' values in $(i+1)th$ layer, which is:

$$c_{i+1} = \sum_{j=1}^{2^i} n_i^j \quad (24)$$

In the tree-based decomposing algorithm, the decomposing rule is shown in (22), i.e., the sum of child-nodes' values should equal to the parent-node's value minus 1. The process of the proposed tree-based decomposing algorithm can be expressed as follows. Firstly, choosing one node (node $i$) as the starting node, and comparing the utility of this node with all the other nodes, so the comparison times is $m$-1, where $m$ is the number of nodes; assuming that the order of node $i$ is $k$. Secondly, based on Corollary 1, there will have $k$-1 nodes whose utilities are smaller than node $i$ and $m$-$k$ nodes whose utilities are larger than node $i$; so, in layer 2, $n_1^1 = k - 1$ and $n_1^2 = m - k$. In this layer, $n_1^1 + n_1^2 = m - 1$. Finally, the process of step 1 and step 2 will be repeated by all the other nodes in the decomposing tree until the value of the node is 0 or 1; moreover, all the values of the nodes need to satisfy the requirement of (22). Based on this process, the comparison times in $l$PMADM can be calculated, which is shown in (23). ∎

Based on Corollary 8 and Corollary 7, the lower bound and higher bound of comparison time can be calculated as $\sum_{i=1}^{\lfloor \log_2 m \rfloor} \sum_{j=1}^{2^i} n_i^j$ and $\sum_{i=1}^{(m-1)} \sum_{j=1}^{2^i} n_i^j$, respectively. Therefore, we have Corollary 9 as follows.

**Corollary 9.** The lower bound of the comparison time can be calculated as:

$$\sum_{i=1}^{\lfloor \log_2 m \rfloor} \sum_{j=1}^{2^i} n_i^j$$
$$= (\lfloor \log_2 m \rfloor - 1)(m-1)$$
$$+ (m - 2^{\lfloor \log_2 m \rfloor} + 1) - (2^{\lfloor \log_2 m \rfloor - 1} - 2) \quad (25)$$

The higher bound of comparison time can be calculated as:

$$\sum_{i=1}^{(m-1)} \sum_{j=1}^{2^i} n_i^j = \frac{m(m-1)}{2} \quad (26)$$

*Proof.* The higher bound of comparison time is obviously. When the starting node is the smallest or the largest, the comparison time is the largest, which is $\frac{m(m-1)}{2}$.

Suppose the value in layer 0 is $m$. Based on the optimal decomposing, in layer 1, we choose one starting node and compare this node with all the other nodes, so the comparison times in layer 1 is $m - 1$. In layer 2, two starting nodes are chosen, we compare these two nodes with the respected nodes, so the comparison times is $m - 1 - 2$. Based on this analysis, the value in layer 3 is $m - 1 - 2 - 2^2$, and so on. This process will be repeated until the node's value equals to 0 or 1. Therefore, the sum of the nodes' values in $ith$ layer is:

$$n_i = \sum_{j=1}^{2^i} n_i^j = m - 1 - 2 - 2^2 - \cdots - 2^{i-1}$$
$$= m - 2^i + 1 \quad (27)$$

According to (27), the sum of all the nodes' value in the decomposing tree can be calculated as:

$$\sum_{i=1}^{\lfloor \log_2 m \rfloor} \sum_{j=1}^{2^i} n_i^j$$
$$= m - 1 + m - 1 - 2 + m - 1 - 2 - 2^2$$
$$+ \cdots + (m - 1 - \cdots - 2^{\lfloor \log_2 m \rfloor - 1}) + m - 2^{\lfloor \log_2 m \rfloor} + 1$$
$$= (\lfloor \log_2 m \rfloor - 1)(m-1)$$
$$+ (m - 2^{\lfloor \log_2 m \rfloor} + 1) - (2^{\lfloor \log_2 m \rfloor - 1} - 2) \quad (28)$$

The value shown in (28) equals to the comparison times of $l$PMADM algorithm. Therefore, Corollary 9 holds. ∎

Moreover, based on the tree-based decomposing algorithm, the comparison times in the higher layer is larger than that in the lower layer, which can be got from (22) and (23). Therefore, for the optimal decomposing, the fewer layers, the fewer comparisons, i.e., if the value of $l$ is the smallest, the value of $m$ is also the smallest. When we get the numbers of divisions and the number of comparisons, we can analyze the computational complexity of the $l$PMADM algorithm.

*Computational complexity.* The computational complexity of $l$PMADM is shown in Corollary 10.

**Corollary 10.** The lower-bound of the computational complexity of $l$PMADM is $O[nm \log(m)]$; the higher-bound of the computational complexity of $l$PMADM is $O[nm^2]$.

*Proof.* Based on Corollary 9, the lower-bound and the higher-bound of the comparison times are $(\lfloor \log_2 m \rfloor - 1)(m-1) + (m - 2^{\lfloor \log_2 m \rfloor} + 1) - (2^{\lfloor \log_2 m \rfloor - 1} - 2)$ and $\frac{m(m-1)}{2}$, respectively. Therefore, the lower-bound and the higher-bound computational complexity of the comparison time are $O[m \log(m)]$ and $O[m^2]$. However, in the $l$PMADM algorithm, for each comparison, the utility of each node needs to be calculated. Since there are $n$ parameters in each node, the lower-bound and the higher-bound of the computational complexity of $l$PMADM algorithm will be $O[nm \log(m)]$ and $O[nm^2]$, respectively. ∎

*C. Fault Tolerance*

In Corollary 5, we conclude that when $m$ is odd, then the starting node is $k = \frac{m+1}{2}$; if $m$ is even, then $k = \frac{m}{2}$ or $k = \frac{m+2}{2}$. When the staring node satisfies the requirements of Corollary 5, the comparison time is the smallest. However, in practice, the value of $k$ may not equal to the values shown in Corollary 5. Even we propose the low-complexity PMADM algorithm in Section IV.A, it still cannot guarantee that the chosen starting node satisfies the requirements in Corollary 5. When the value of $k$ does not equal to these values, the comparison times may increase. However, in the following, we will prove that our proposed tree-based decomposing algorithm can tolerant this error to some extent. We call this the fault tolerance.

**Definition 3.** The fault tolerance of the tree-based decomposing algorithm is defined as the number of the division pairs whose comparison times are the same as the optimal division (i.e., the division which can satisfy the requirements in Corollary 5), denoted as $\delta$.

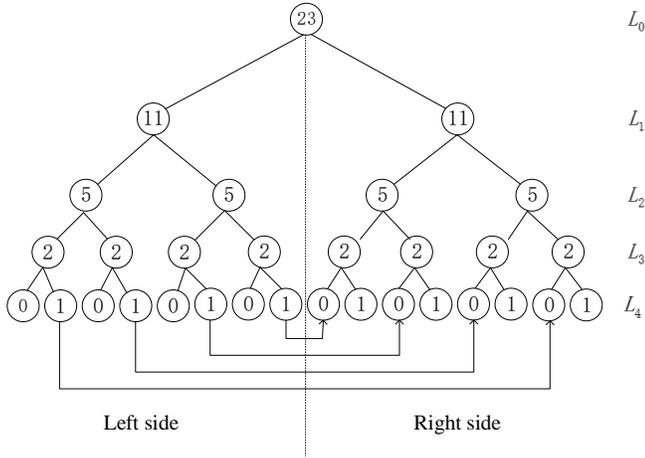

Fig. 3 An example of optimal tree-based decomposing

Based on Definition 3, we can calculate the fault tolerance of the tree-based decomposing algorithm. Before that, we first introduce how to calculate the sum of nodes' values in the final layer. Suppose that the node's value in layer 0 is $m$. If the nodes' values in layer 1 are $n_1^1$ and $n_1^2$, then based on Corollary 6, the numbers of layers of these two nodes can be calculated, which are $l_1^1$ and $l_1^2$. Additionally, the sum of the nodes' values in the final layers of $n_1^1$ and $n_1^2$ also can be calculated based on Corollary 9, which are $\varepsilon_1^1$ and $\varepsilon_1^2$. For instance, as shown in Fig.3, $\varepsilon_1^1$ is the sum of nodes' values in the final layers of the left side; $\varepsilon_1^2$ is the sum of nodes' values in the final layers of the right side. So, the number of layers of $m$ can be calculated as: $l_0 = \max\{l_1^1, l_1^2\}$. However, the calculation of the sum of the nodes' values in the last layer of $m$ is a little complicated. The calculation needs to be divided into three aspects:

(1) if $l_1^1 = l_1^2$, then the sum can be calculated as (shown in Fig.3):
$$\varepsilon_0 = \varepsilon_1^1 + \varepsilon_1^2 \tag{29}$$

(2) if $l_1^1 > l_1^2$, then the sum can be calculated as (shown in Fig.2):
$$\varepsilon_0 = \varepsilon_1^1 \tag{30}$$

(3) if $l_1^1 < l_1^2$, then the sum can be calculated as (same as Fig.2):
$$\varepsilon_0 = \varepsilon_1^2 \tag{31}$$

where the $\varepsilon_1^1$ and $\varepsilon_1^2$ can be calculated based on the conclusion in Corollary 9.

**Corollary 11.** The fault tolerance of the tree-based decomposing algorithm is $\delta = \min\{\varepsilon_{max}, 2^{\lfloor \log_2 n_{max} \rfloor} - \varepsilon_{min}\} + 1$ or 1, where $\varepsilon_{min} = \min\{\varepsilon_1^1, \varepsilon_1^2\}$, $\varepsilon_{max} = \max\{\varepsilon_1^1, \varepsilon_1^2\}$, and $n_{max} = \max\{n_1^1, n_1^2\}$.

*Proof.* As shown in Fig. 3, when the decomposing is optimal, for the left side of the decomposing tree, the sum of the nodes' values in the final layer is decided by $n_1^1$, which is $\varepsilon_1^1$. Similarly, for the right side, the sum of the nodes' values in the final layer is decided by $n_1^2$, which is $\varepsilon_1^2$. As shown in Fig.3, the number of "0" and the number of "1" in the final layers of the left side and right side can be calculated based on $n_1^1$ and $n_1^2$, respectively.

Suppose the optimal decomposing in layer 1 are $n_1^1$ and $n_1^2$, i.e., the $n_1^1$ and $n_1^2$ satisfy the requirements of Corollary 5, then the number of layers and the sum of the nodes' values in the last layer of $m$ can be calculated, which are $l^*$ and $f^*$, respectively. Assuming that another decomposing in layer 1 is $n_1^{1-1}$ and $n_1^{2-1}$, if the number of layers and the comparison times are the same as the optimal decomposing, first, $n_1^* = \max\{n_1^1, n_1^2\}$ and $n_1^{1-*} = \max\{n_1^{1-1}, n_1^{2-1}\}$ should in the same range, i.e., $l_1^* = l_1^{1-*}$; second, $f_0^*$ and $f_0^{1-*}$ are equally; this is because when $l_1^* = l_1^{1-*}$, the comparison times from layer 1 to $l_1^* - 1$ are the same in this two different decomposing. So, the sum of the nodes' values in the last layers of the left side and right side can be calculated based on (29), (30), and (31), where $\varepsilon_1^1 = n_1^1 + 1 - 2^{\lfloor \log_2 n_1^1 \rfloor}$ and $\varepsilon_1^2 = n_1^2 + 1 - 2^{\lfloor \log_2 n_1^2 \rfloor}$.

In the optimal decomposing, there are $n_1^1 = n_1^2$ (when $m$ is odd) or $n_1^1 = n_1^2 + 1$[1] (when $m$ is even). If $n_1^1$ and $n_1^2$ are in the same range, i.e., $n_1^1$ and $n_1^2$ satisfy $2^{\lfloor \log_2 n \rfloor} \leq n < 2^{\lfloor \log_2 n \rfloor+1}$, then $l_1^1 = l_1^2$ and $\varepsilon_1^1 = \varepsilon_1^2$ or $\varepsilon_1^1 = \varepsilon_1^2 + 1$. If $l_1^1 = l_1^2$, $\varepsilon = \varepsilon_1^1 + \varepsilon_1^2$. If $n_1^1$ and $n_1^2$ are not the optimal values, such as $n_1^{1-1} = n_1^1 - 1$ and $n_1^{2-1} = n_1^2 + 1$, then $\varepsilon_1^{1-1} = \varepsilon_1^1 - 1$ and $\varepsilon_1^{2-1} = \varepsilon_1^2 + 1$. This equals to $\varepsilon = \varepsilon_1^1 - 1 + \varepsilon_1^2 + 1 = \varepsilon_1^1 + \varepsilon_1^2$. This gives us an inspiration that we can transfer the "1" from the left side/right side to the right side/left side while maintains the total number of "1" in the final layer is constant. When the total number of "1" in the final layer is constant, the comparison time is also constant. This equals the optimal comparison time. An example is shown in Fig.3. So, in the decomposing tree, when all the nodes in the final layer of one side are all 0 or all the nodes in the final layer of one side are all 1, this process is stopped. Because any further transferring will cause the increase of the number of layers and the comparison times. We can conclude this conclusion is because when the values in $n_1^1$ increase or decrease 1, the number of nodes in the final layer whose values equal to 1 will increase or decrease 1 correspondingly. In this scenario, on one hand, the maximum times of transferring is the maximum number of "1" in left side and right side; on the other hand, the transferring should not overflow, i.e., the input side of the transferring should have enough "0" to hold these transferred "1". Therefore, the fault tolerance in the tree-based decomposing algorithm is $\delta = \min\{\varepsilon_{max}, 2^{\lfloor \log_2 n_{max} \rfloor} - \varepsilon_{min}\} + 1$, $\varepsilon_{min} = \min\{\varepsilon_1^1, \varepsilon_1^2\}$, $\varepsilon_{max} = \max\{\varepsilon_1^1, \varepsilon_1^2\}$, and $n_{max} = \max\{n_1^1, n_1^2\}$.

---

[1] Since $n_1^1 = n_1^2 + 1$ and $n_1^2 = n_1^1 + 1$ are symmetrically, we omit the discussion when $n_1^2 = n_1^1 + 1$ in this paper.

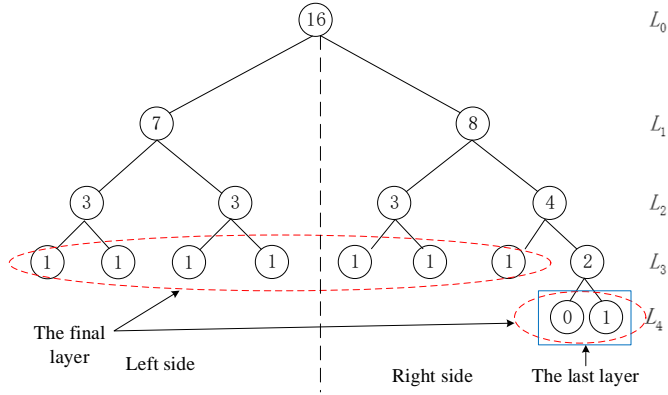

Fig.4 An example of the decomposing tree

When $n_1^1$ and $n_1^2$ are not in the same range, i.e., $n_1^1$ and $n_1^2$ satisfy $n_1^1 < 2^{\lfloor \log_2 n_1^2 \rfloor} \leq n_1^2 < 2^{\lfloor \log_2 n_1^2 \rfloor + 1}$, then $l_1^1 = l_1^2 - 1$. This is shown in Fig.4. In this scenario, since layer 4 is empty on the left side, the transferring cannot be achieved. Therefore, the number of decomposing is 1 in this scenario. ∎

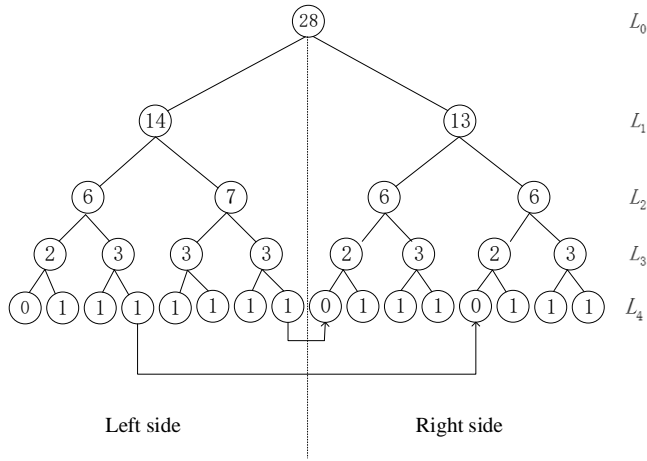

Fig.5 An example of the decomposing tree

*An example of Corollary 14.* As shown in Fig.5, on the left side $\varepsilon_1^1 = 7$, on the right side $\varepsilon_1^2 = 6$. So, for the left side, 7 nodes can be transferred to the right side; for the right side, 6 nodes can be transferred to the left side. However, there is only one "0" on the left side and two "0" on the right side. Therefore, at most two "1" can be transferred from the left side to the right side. Therefore, there will 3 kinds of decomposing which have the same comparison times that equal to the optimal value. This represents that the fault tolerance of this scenario is 3. Based on Corollary 11, we have the conclusions as follows.

**Corollary 12.** For the tree-based decomposing algorithm, when the nodes in the same range, i.e., $2^{\lfloor \log_2 m \rfloor} \leq m < 2^{\lfloor \log_2 m \rfloor + 1}$, the node which has the largest fault tolerance can be calculated as: if $\tau = 2^{\lfloor \log_2 m \rfloor} + \frac{2^{\lfloor \log_2 m \rfloor + 1} - 2^{\lfloor \log_2 m \rfloor}}{2}$ is odd, then $\tau$ has the largest fault tolerance in this range; if $\tau = 2^{\lfloor \log_2 m \rfloor} + \frac{2^{\lfloor \log_2 m \rfloor + 1} - 2^{\lfloor \log_2 m \rfloor}}{2}$ is even, then $\tau - 1$ has the largest fault tolerance.

*Proof.* For the nodes which are in the same range, in their decomposing tree, the number of layers is consistently based on the optimal decomposing, which is presented in Corollary 6. In the decomposing tree, for the values of nodes in the same range, the number of nodes in the final layer may different. However, the number of layers when the nodes' values are in the same range is the same. For instance, when the value of $n_0$ is 17, based on the decomposing tree, the nodes in the final layer is 4, which is shown in Fig.6. This is different from that shown in Fig.3. This is because some nodes in the higher-layer are already equal to 0 or 1, so in the final layer, these nodes do not need to be decomposed further. For instance, in Fig.6, the values in L3 and L4.

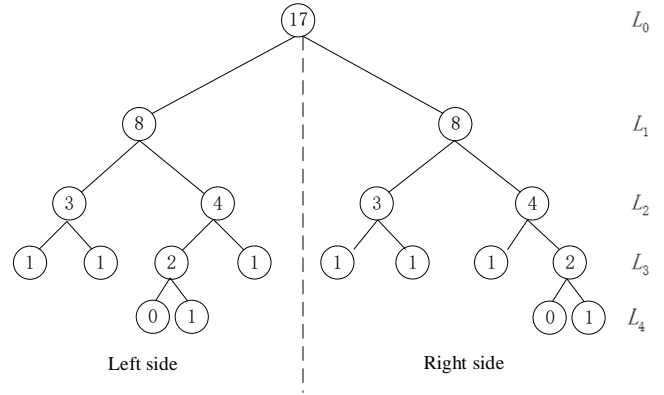

Fig.6 An example of the decomposing tree

The maximum number of nodes in the final layer is $2^{\lfloor \log_2 m \rfloor}$. Based on our conclusion in Corollary 11, if the fault tolerance is the maximum, two conditions should be satisfied: first, the number of nodes in the final layer should equal to the maximum value, i.e., $2^{\lfloor \log_2 m \rfloor}$; second, the $\delta = \min\{\varepsilon_{max}, 2^{\lfloor \log_2 n_{max} \rfloor} - \varepsilon_{min}\} + 1$ in Corollary 11 should get the minimum value. Moreover, as shown in Fig.2, two nodes which have the same parent-node in the final layer cannot equal to 0 simultaneously. Because when these two nodes are equal to 0, it means that the value of their parent-node must equal to 0 or 1, which cannot be decomposed further. Based on this analysis, the node which has the highest fault tolerance should satisfy two conditions: first, the number of nodes in the final layer should equal to $2^{\lfloor \log_2 m \rfloor}$ and the number of nodes in the right side equals to the left side; second, in each side, there should be half nodes equal to 0 and half nodes equal to 1. When the second condition is satisfied, we have $n_1^1 = n_1^2$. Therefore, if $\tau = 2^{\lfloor \log_2 m \rfloor} + \frac{2^{\lfloor \log_2 m \rfloor + 1} - 2^{\lfloor \log_2 m \rfloor}}{2}$ is odd, then $\tau$ has the largest fault tolerance in this range; if $\tau = 2^{\lfloor \log_2 m \rfloor} + \frac{2^{\lfloor \log_2 m \rfloor + 1} - 2^{\lfloor \log_2 m \rfloor}}{2}$ is even, then the value $\tau - 1$ has the largest fault tolerance. Therefore, Corollary 12 holds. ∎

However, the conclusions in the corollaries mentioned above hold only when the rest nodes except the nodes in layer 1 are decomposed optimally. However, this is not always hold in practice. So, we give the Corollary 13 as follows.

**Corollary 13.** The fault tolerance of the whole decomposing tree can be calculated as:

$$\delta = \prod_{i=1}^{\lfloor \log_2 m \rfloor} \sum_{j=1}^{j=2^i} \delta_i^j \quad (32)$$

where $\delta_i^j$ is the $j$th fault tolerance of layer $i$ and $i \leq \lfloor \log_2 m \rfloor$.

*Proof.* This conclusion is obvious, science there is more than one layer in the decomposing tree and the fault in each layer can make the decomposing does not equal to the optimal value. However, the problem is that, for layer 1, there is only one fault tolerance; for the other layers, this value will larger than 1 because the decomposing is more than once. For instance, as the decomposing tree shown in Fig.3, from layer 1 to layer 2, there will be two different fault tolerances, because there are two decomposing from layer 1 to layer 2; from layer 2 to layer 3, there will be four different fault tolerances, and so on. So, the fault tolerance in $ith$ layer can be expressed as:

$$\delta_i = \sum_{j=1}^{j=2^i} \delta_i^j \quad (33)$$

where $i \leq \lfloor \log_2 m \rfloor$. Additionally, based on the permutation and combination theory, the fault tolerance which takes all the layers in the decomposing tree into account can be calculated as $\delta = \prod_{i=1}^{\lfloor \log_2 m \rfloor} \sum_{j=1}^{j=2^i} \delta_i^j$. ∎

**Corollary 14.** To different ranges, for the tree-based decomposing algorithm, the probability that the decomposing is optimal (shorted as the probability of optimal) decreases with the increasing of $m$, i.e., if $m$ is large, then under the optimal decomposing, the probability of optimal under the larger $m$ is smaller than the smaller $m$.

*Proof.* From Corollary 11, the fault tolerance of the tree-based decomposing algorithm is $N = \min\{\varepsilon_{max}, 2^{\lfloor \log_2 n_{max} \rfloor} - \varepsilon_{min}\} + 1$ or 1. According to Corollary 9, the sum of the nodes' values in the final layer is $m - 2^{\lfloor \log_2 m \rfloor} + 1$. Moreover, the total number of decomposing pairs is:

$$M = \left\lfloor \frac{m}{2} \right\rfloor \quad (34)$$

So, based on the conclusion in Corollary 11, the probability of optimal can be calculated as:

$$\rho = \frac{N}{M} = \frac{n_1^1 - 2^{\lfloor \log_2 n_1^1 \rfloor} + 1}{\left\lfloor \frac{m}{2} \right\rfloor} \quad (35)$$

or

$$\rho = \frac{N}{M} = \frac{2^{\lfloor \log_2 n_1^2 \rfloor} - \left(n_1^2 - 2^{\lfloor \log_2 n_1^2 \rfloor} + 1\right)}{\left\lfloor \frac{m}{2} \right\rfloor} = \frac{2^{\lfloor \log_2 n_1^2 \rfloor + 1} - n_1^2 - 1}{\left\lfloor \frac{m}{2} \right\rfloor} \quad (36)$$

or

$$\rho = \frac{N}{M} = \frac{1}{\left\lfloor \frac{m}{2} \right\rfloor} \quad (37)$$

where $n_1^1 + n_1^2 = m - 1$. So, the (36) can be rewritten as:

$$\rho = \frac{N}{M} = \frac{2^{\lfloor \log_2 (m-1-n_1^1) \rfloor} - \left(m-1-n_1^1 - 2^{\lfloor \log_2 (m-1-n_1^1) \rfloor} + 1\right)}{\left\lfloor \frac{m}{2} \right\rfloor}$$

$$= \frac{2^{\lfloor \log_2 n_1^2 \rfloor + 1} + n_1^1 - m}{\left\lfloor \frac{m}{2} \right\rfloor} \quad (38)$$

With the increasing of $m$, the $n_1^1$ and $n_1^2$ have the same increasing rate as $m$. Moreover, we assume that in the optimal decomposing, $\left\lfloor \frac{m}{2} \right\rfloor = n_1^1$ and $n_1^2 = n_1^1$ or $n_1^2 = n_1^1 + 1$; then with the increasing of $m$, the values of (35), (36), and (37) decrease. This means that the probability of optimal in different ranges decreases. Therefore, Corollary 14 holds. ∎

## V. THE PROPERTIES OF PMADM AND $l$PMADM[2]

### A. The effectiveness of PAMDM algorithm

**Corollary 15.** The proposed PMADM algorithm can change the sequence of the node, i.e., the proposed PMADM algorithm can give different decisions from the traditional MADM algorithm.

*Proof.* For the MADM algorithm, assuming that the weights of parameters are $\omega_A$, $\omega_B$, $\omega_C$, and $\omega_D$; so, $\omega_A + \omega_B + \omega_C + \omega_D = 1$. Based on (1), the utility of $U_1$ and $U_2$ are:

$$U_1 = \omega_A P_A^1 + \omega_B P_B^1 + \omega_C P_C^1 + \omega_D P_D^1 \quad (39)$$

$$U_2 = \omega_A P_A^2 + \omega_B P_B^2 + \omega_C P_C^2 + \omega_D P_D^2 \quad (40)$$

So, $U_1 - U_2$ can be calculated as:

$$U_{12} = U_1 - U_2 = \omega_A P_A^{12} + \omega_B P_B^{12} + \omega_C P_C^{12} + \omega_D P_D^{12} \quad (41)$$

For the PMADM algorithm, for N1 and N2, the weights of parameters are $\omega_A^{12}$, $\omega_B^{12}$, $\omega_C^{12}$, and $\omega_D^{12}$; so, $\omega_A^{12} + \omega_B^{12} + \omega_C^{12} + \omega_D^{12} = 1$. Based on PMADM algorithm, the utilities of $U_{12}^1$ and $U_{12}^2$ are:

$$U_{12}^1 = \omega_A^{12} P_A^1 + \omega_B^{12} P_B^1 + \omega_C^{12} P_C^1 + \omega_D^{12} P_D^1 \quad (42)$$

$$U_{12}^2 = \omega_A^{12} P_A^2 + \omega_B^{12} P_B^2 + \omega_C^{12} P_C^2 + \omega_D^{12} P_D^2 \quad (43)$$

Thus, $U_{12}^1 - U_{12}^2$ can be calculated as:

$$U_{12}^{12} = U_{12}^1 - U_{12}^2$$
$$= \omega_A^{12} P_A^{12} + \omega_B^{12} P_B^{12} + \omega_C^{12} P_C^{12} + \omega_D^{12} P_D^{12} \quad (44)$$

Therefore, we need to prove that if $U_{12} > 0$, then $U_{12}^{12}$ may be smaller than 0. According to (41) and assume that $P_B^{12}$, $P_C^{12}$, and $P_D^{12}$ are smaller than 0, we have:

$$U_{12} = \omega_A P_A^{12} + \omega_B P_B^{12} + \omega_C P_C^{12} + \omega_D P_D^{12}$$
$$= \omega_A P_A^{12} - (\omega_B P_B^{12} + \omega_C P_C^{12} + \omega_D P_D^{12})$$
$$< \omega_A P_A^{12} - M \cdot (\omega_B + \omega_C + \omega_D)$$
$$= \omega_A P_A^{12} - M \cdot (1 - \omega_A) > 0$$
$$= \omega_A (P_A^{12} + M) - M > 0 \quad (45)$$

where $M = \max\{|P_B^{12}|, |P_C^{12}|, |P_D^{12}|\}$. Similarly, in the PMADM algorithm, based on (44) and the fact that in these two different algorithms, the value of $P_A^{12}$, $P_B^{12}$, $P_C^{12}$, and $P_D^{12}$ are the same; only the weights change. So, if the PMADM algorithm can

---

[2]Since the PMADM and $l$PMADM have the same properties discussed in this section, so we only show the properties of PMADM algorithm. The properties of $l$PMADM algorithm are the same as those shown in this section.

change the sequences of nodes, the following condition should be satisfied:

$$U_{12} = \omega_A^{12} P_A^{12} + \omega_B^{12} P_B^{12} + \omega_C^{12} P_C^{12} + \omega_D^{12} P_D^{12}$$
$$= \omega_A^{12} P_A^{12} - (\omega_B^{12} P_B^{12} + \omega_C^{12} P_C^{12} + \omega_D^{12} P_D^{12})$$
$$< \omega_A^{12} P_A^{12} - M \cdot (\omega_B^{12} + \omega_C^{12} + \omega_D^{12})$$
$$= \omega_A^{12} P_A^{12} - M \cdot (1 - \omega_A^{12}) < 0$$
$$= \omega_A^{12}(P_A^{12} + M) - M < 0 \quad (46)$$

Thus, if (46) holds, the $\omega_A^{12}$ should be smaller than $\omega_A$. As introduced in Section III, the $\omega_A$ relates to the variance of all the values of $P_A$. We assume the values of $P_A$ are $P_A^1, P_A^2, P_A^3, P_A^4$, and $P_A^5$. The $\omega_A^{12}$ only relates to the variance of $P_A^1$ and $P_A^2$. So, if $\omega_A^{12} < \omega_A$, it means that $P_A^3, P_A^4$, and $P_A^5$ vary greatly than $P_A^1$ and $P_A^2$. This is possible. Thus, the Corollary 15 holds. ∎

The Corollary 15 illustrates that the PMADM algorithm can change the sequence of node calculated by the traditional MADM algorithm. This means that the PAMDM algorithm is effective. Moreover, as shown in (46), the smaller of $\omega_A^{12}$ than $\omega_A$, the more possible of PAMDM algorithm change the sequence calculated by the traditional MADM algorithm.

### B. The fairness and accuracy

**Corollary 16.** The PMADM algorithm is much more accurate and fairer than the traditional MADM algorithm.

*Proof.* As discussed in the Corollary 15, if $\omega_A^{12} < \omega_A$, then the proposed PMADM algorithm can change the sequences that calculated by the traditional MADM algorithm. Moreover, $\omega_A$ relates to the variance of all the values of $P_A$, i.e., $P_A^1, P_A^2, P_A^3, P_A^4$, and $P_A^5$. The $\omega_A^{12}$ only relates to the variance of $P_A^1$ and $P_A^2$. So, $\omega_A^{12} < \omega_A$ means that $P_A^3, P_A^4$, and $P_A^5$ vary greatly than $P_A^1$ and $P_A^2$. In this scenario, if we still use the traditional MADM algorithm to calculate the utility of each node, it will be unfair because the $\omega_A$ covers up the real relationship between nodes. However, for the PMADM algorithm, $\omega_A^{12} < \omega_A$ means that the real variance between $P_A^1$ and $P_A^2$ is small. So, if we use $\omega_A$ as the weight, it is unfair. The PMADM algorithm is fairer than the traditional MADM algorithm.

For the traditional MADM algorithms, as shown in Section V.A, if the value of $P_A^5$ is much larger than that of $P_A^1, P_A^2, P_A^3$, and $P_A^4$, then the weight of $P_A$ (i.e., $\omega_A$) will be very high. Assuming that the values of $P_A^1, P_A^2, P_A^3$, and $P_A^4$ are similar, and $\omega_A$ is larger than $\omega_B, \omega_C$, and $\omega_D$. So, if $P_A^5$ is too large, it will have two effects on the final decision. On the one hand, because the values of $P_A^1, P_A^2, P_A^3$, and $P_A^4$ are similar to each other, the variance of these four parameters are small, which is much smaller than that of $\omega_A$. So, in the traditional MADM algorithm, $\omega_A$ cannot reflect the real relationship between $P_A^1, P_A^2, P_A^3$, and $P_A^4$. This will affect the final decision of the algorithm. On the other hand, since $\omega_A + \omega_B + \omega_C + \omega_D = 1$, so if $\omega_A$ increases, the value of $\omega_B, \omega_C$, and $\omega_D$ will be decreased. This makes the final decision unfair further. This is because the parameters affect each other greatly in the traditional MADM algorithm. When we compare the utility of node $i$ and node $j$, the parameters in other nodes can influence the results by the traditional MADM algorithm. We call this the parameter pollution.

However, in the PMADM algorithm, we calculate the utility between only two nodes at each time. This can isolate the parameter pollution. For instance, if we compute the utilities of node $i$ and node $j$, then only the parameters of node $i$ and node $j$ can affect each other. Even the parameters in other nodes vary greatly, they cannot influence the calculation result of node $i$ and node $j$. This is one of the main advantages of PMADM, which is much more accurate than MADM. Therefore, Corollary 16 holds. ∎

### C. The changing threshold

As introduced in Corollary 15, the proposed PMADM algorithm can change the sequence of node calculated by the traditional MADM algorithm. In the following, we will give the threshold of changing the sequences of nodes.

**Corollary 17.** The threshold of the node utility which can make the node sequence variant is $\Delta u$; the lower-bound of $\Delta u$ is $|\Delta u_{ij}| > |u_{ij}|$. Moreover, if $U_i > U_j$ and the sequence changes based on the PMADM algorithm, then the PMADM algorithm is a benefit to N$j$.

*Proof.* For the traditional MADM algorithm, the utility of N1 and N2 can be calculated based on (1), which are $U_1$ and $U_2$, respectively. Moreover, we assume that $U_1 > U_2$. For the PMADM algorithm, the utility of N1 and N2 can be computed based on (2), which are $U_1^*$ and $U_2^*$, respectively. For these four utilities, their relationship can be expressed as:

$$U_1^* = U_1 + \Delta U_1 \quad (47)$$
$$U_2^* = U_2 + \Delta U_2 \quad (48)$$

So, if the PMADM can make the sequences of nodes that are decided by the traditional MADM algorithm changeable, then the following condition should be satisfied.

$$(U_1 - U_2)(U_1^* - U_2^*) < 0 \quad (49)$$

Based on the (47) and (48), the (49) can be rewritten as:

$$(U_1 - U_2)(U_1 + \Delta U_1 - U_2 - \Delta U_2) < 0$$
$$= (U_1 - U_2)(U_1 - U_2 + \Delta U_1 - \Delta U_2) < 0 \quad (50)$$

Since $U_1 - U_2 = U_{12}$ and $\Delta U_1 - \Delta U_2 = \Delta U_{12}$, we have:

$$U_{12}(U_{12} + \Delta U_{12}) < 0 \quad (51)$$

Since $U_1 > U_2$, $U_{12} > 0$. However, $\Delta U_1$ and $\Delta U_2$ may be larger or smaller than 0, so $\Delta U_{12}$ may also be larger or smaller than 0. Therefore, if $\Delta U_{12} > 0$, then (51) does not hold. If (51) holds, there must be $\Delta U_{12} < 0$, which equals to:

$$U_{12} < -\Delta U_{12} \quad (52)$$

this equals to $|U_{12}| < |\Delta U_{12}|$.

According to the analysis above, if $U_1 > U_2$, then there must exist $\Delta U_{12} < 0$, i.e., $\Delta U_1 - \Delta U_2 < 0$. So, if $\Delta U_1 > 0$, then $\Delta U_2 > 0$ and $\Delta U_2 > \Delta U_1$. This demonstrates that based on these two algorithms, the weights of the parameters in N2 change more greatly than that in N1. So, this means that the PAMDM algorithm is benefit to N2 rather than N1. If $\Delta U_1 < 0$, then if $\Delta U_2 > 0$, then $\Delta U_2$ may be larger or smaller than $|\Delta U_1|$; if $\Delta U_2 < 0$, then there must be $|\Delta U_1| > |\Delta U_2|$. This also illustrates that the PAMDM algorithm is a benefit to N2. Therefore, if $U_1 > U_2$ and $U_1^* < U_2^*$ holds, the PMADM algorithm is a benefit to N2. Thus, the Corollary 17 holds. ∎

The Corollary 17 demonstrates that if the sequences of N$i$ and N$j$ can be changed by the PMADM algorithm, the difference of the utilities between these two nodes should be larger than that calculated by the traditional MADM algorithm.

### D. The sensibility of PMADM and MADM

For different decision making algorithms, the sensibility of the algorithm can be defined as follows.

**Definition 4.** The sensibility of MADM algorithm is that for two MADM algorithms and two nodes: node $i$ and node $j$, when the same parameter changes in these two nodes, for these two algorithms, the threshold of the parameter changing can change the sequences of node $i$ and node $j$.

Based on the conclusion that the parameters whose variance is large have a great effect on the routing performances, we conclude that a good MADM algorithm should be sensitive to the parameters whose values deviate from the average value seriously; however, for the parameters whose value is close to the average value, the algorithm should not be sensitive. In the following corollary, we will show that the PMADM algorithm has these properties.

**Corollary 18.** When the parameters whose values deviate from the average value seriously, the PMADM is much more sensitive than the MADM; when the parameters whose value approximates to the average value, the MADM is more sensitive than the PMADM.

*Proof.* For the traditional MADM algorithm, the utilities of N1 and N2 can be calculated based on (39) and (40), respectively. To the PMADM algorithm, the utilities of N1 and N2 can be calculated based on (42) and (43), separately.

So, the sensibility of the traditional MADM algorithm and the PMADM algorithm when $P_A$ changes can be calculated as:

$$\omega_A \cdot \Delta P_A^1 = -(\omega_A P_A^{12} + \omega_B P_B^{12} + \omega_C P_C^{12} + \omega_D P_D^{12}) \quad (53)$$

$$\omega_A^{12} \cdot \Delta P_A^{1*} = -(\omega_A^{12} P_A^{12} + \omega_B^{12} P_B^{12} + \omega_C^{12} P_C^{12} + \omega_D^{12} P_D^{12}) \quad (54)$$

This is because of $\overline{P_A^1} = P_A^1 + \Delta P_A^1$, $\overline{P_A^{12}} = P_A^1 + \Delta P_A^1 - P_A^1 = P_A^{12} + \Delta P_A^1$. Moreover, $\omega_A + \omega_B + \omega_C + \omega_D = 1$ and $\omega_A^{12} + \omega_B^{12} + \omega_C^{12} + \omega_D^{12} = 1$.

The (53) and (54) can be rewritten as:

$$\Delta P_A^1 = -\left(P_A^{12} + \frac{\omega_B}{\omega_A} P_B^{12} + \frac{\omega_C}{\omega_A} P_C^{12} + \frac{\omega_D}{\omega_A} P_D^{12}\right) \quad (55)$$

$$\Delta P_A^{1*} = -\left(P_A^{12} + \frac{\omega_B^{12}}{\omega_A^{12}} P_B^{12} + \frac{\omega_C^{12}}{\omega_A^{12}} P_C^{12} + \frac{\omega_D^{12}}{\omega_A^{12}} P_D^{12}\right) \quad (56)$$

Since $\omega_A + \omega_B + \omega_C + \omega_D = 1$ and $\omega_A^{12} + \omega_B^{12} + \omega_C^{12} + \omega_D^{12} = 1$, and we assume that $P_B^{12}$ is the largest in $P_B^{12}$, $P_C^{12}$, and $P_D^{12}$, then (55) and (56) equal to:

$$\Delta P_A^1 \geq -\left(P_A^{12} + \frac{1-\omega_A}{\omega_A} P_B^{12}\right) \quad (57)$$

$$\Delta P_A^{1*} \geq -\left(P_A^{12} + \frac{1-\omega_A^{12}}{\omega_A^{12}} P_B^{12}\right) \quad (58)$$

So, if $\omega_A < \omega_A^{12}$, $|\Delta P_A^1| > |\Delta P_A^{1*}|$ holds; if $\omega_A > \omega_A^{12}$, $|\Delta P_A^1| < |\Delta P_A^{1*}|$ holds. These conclusions mean that if $\omega_A^{12}$ is much larger than $\omega_A$, i.e., the values of $P_A$ vary greatly, then the changing of this parameter is sensitive in PMADM algorithm; however, when $\omega_A^{12}$ is much smaller than $\omega_A$, i.e., the value of $P_A$ does not change greatly, then the variation of this parameter is not sensitive in PMADM algorithm. ∎

This property is important, because in the decision making theory, the parameters which vary greatly have a great effect on the performances of the final decision, so the algorithm should be sensitive to these kinds of parameters. However, the parameters which vary slightly have fewer effects on the final decision, so the algorithm could be insensitive to the changing of these parameters unless the variation is large enough. However, the traditional MADM algorithm does not have this property. Therefore, we claim that our proposed PMADM algorithm is variable sensitivity while the traditional MADM algorithm is not. This is also one of the important advantages over the traditional MADM algorithms.

### E. The Stability of PMADM and MADM

Except for the sensibility, we also explore the stability of the PMADM algorithm and the traditional MADM algorithm.

**Definition 5.** The stability of the algorithm is that when the parameter changes in one node, the relative sequences of the other nodes should not change, only the sequence of this node may change.

Definition 5 means that a good algorithm should not affect other nodes' relative sequences when one node's parameter varies. For instance, there are five nodes: N1, N2, N3, N4, and N5; the sequence of these five nodes is N1>N2>N3>N4>N5. If one of the parameters in N5 changes and makes the utility of N5 larger than N3 and N4, then a stable algorithm should guarantee that the sequence of nodes is N1>N2>N5>N3>N4, i.e., the relative sequences of N1, N2, N3, and N4 should not change.

**Corollary 19.** The PMADM algorithm is much more stable than the previous works, i.e., when $P_i$ changes, only the sequence of N$i$ changes, the relative sequence of the rest nodes are consistent.

*Proof.* For the PMADM algorithm, if one parameter in one node changes, such as $P_1^5$, only the weights of N1 and N5, N2 and N5, N3 and N5, N4 and N5 are affected. The other weights are stable. So, based on (6) and (7), the utilities of $U_{51}$, $U_{52}$, $U_{53}$, and $U_{54}$ are affected; the values of the other utilities are not affected. Therefore, the relative sequences of the other nodes do not change, only the relative sequence of N5 varies.

However, for the traditional MADM algorithm, when the $P_1^5$ changes, the value of $\omega_1$ will be affected, so all nodes' utilities will be affected. Thus, the sequence of these five nodes changes. Therefore, Corollary 19 holds. ∎

This corollary is important. Because for a good MADM algorithm, the nodes' utilities and relative sequences should not change when their parameters are consistent. However, since the weight calculation in the traditional MADM algorithm takes all the parameters into account, it cannot satisfy the requirements of the stability. However, the proposed PMADM algorithm can address this kind of problem perfectly. This property can decrease the computational complexity of the PMADM algorithm. For instance, when $P_1^5$ changes, we only need to re-calculate the relative utilities of N5 to the other nodes. The relative utilities of the other four nodes do not need to be computed, and the relative orders of these four nodes are consistent. However, for the traditional MADM algorithm, when $P_1^5$ changes, all the sequences will be changed, and all the utilities need to be re-calculated. From this point-of-view, the PMADM is much more effective than the traditional MADM algorithm on dealing with the data dynamic.

## VI. SIMULATION

In Section V, the effectiveness, the fairness, the sensibility, and the stability of PMADM are analyzed. So, in this section, we use the routing decision as an example to evaluate the performances of the proposed PMADM algorithm and *l*PMADM algorithm. In this section, we apply these two algorithms and the traditional MADM algorithm into the wireless routing algorithm. The simulation is divided into two different parts: when the parameters vary slightly and when the parameters change greatly. Moreover, we compare the performances of PMADM algorithm and *l*PMADM algorithm in routing decision with the traditional MADM algorithm. However, the PMADM and *l*PMADM algorithm can be used not only in routing decisions but also in other decision-making research areas, such as cloud computing, data center networks, network selection [9], etc. The simulation parameters are shown in Table 3. The variable is the number of nodes.

TABLE 3. SIMULATION PARAMETERS

| simulation parameter | value |
|---|---|
| simulation area | $1000m \times 1000m$ |
| number of nodes | 100, 150, …, 300 |
| maximum transmission range | $250m$ |
| channel data rate | 1Mbps |
| the traffic type | Constant Bit Rate (CBR) |
| packet size | 512bytes |
| beacon interval | $1s$ |
| maximum packet queue length | 50 packets |
| MAC layer | IEEE 802.ll DCF |
| simulation tool | NS2 |

The performance matrixes used in this paper are the transmission delay, the packet delivery ratio, and the network throughput: (1) *packet delivery ratio*: the packet delivery ratio is defined as the ratio of the number of packets received successfully by the destination node to the number of packets generated by the source node [18][19]; (2) *transmission delay*: the transmission delay of the data packet from the source node to the destination node; (3) *network throughput*: the network throughput is the ratio of the total number of packets received successfully by the destination node to the number of packets sent by all the nodes during the simulation [20].

### A. The effectiveness of the PMADM and lPMADM algorithm

In this section, the effectiveness of the PMADM and *l*PMADM algorithm are simulated. In this simulation, the transmission delay, the packet delivery ratio, and the network throughput by using these three algorithms are presented. For each performance matrix, the simulation results are divided into two different parts: when the parameters change slightly and when the parameters vary greatly. The transmission delay is presented in Fig. 7 (in which the parameters vary slightly) and Fig. 8 (in which the parameters vary greatly). From Fig. 7 and Fig. 8, we can conclude that with the increasing of the number of nodes, the transmission delay decreases both in these three algorithms. On the one hand, the performances of transmission delay in PMADM and *l*PMADM algorithm is much better than that in the MADM algorithm. On the other hand, when the network parameters vary slightly, the PMADM has a little better performances than the MADM algorithm; for instance, in Fig. 7, when the number of nodes is 250, the transmission delay of PMADM algorithm is 25%, which is smaller than that of MADM; however, when the parameters change greatly, this value is 40% (shown in Fig. 8). These prove the effectiveness of the PMADM algorithm.

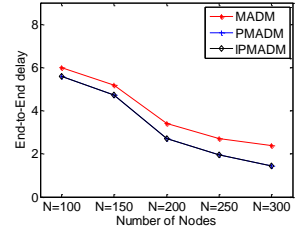 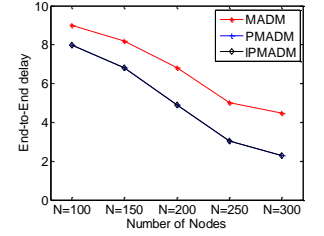

Fig. 7. Transmission delay when the parameters vary slightly  
Fig. 8. Transmission delay when the parameters vary greatly

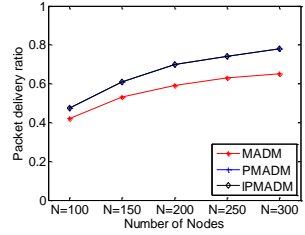 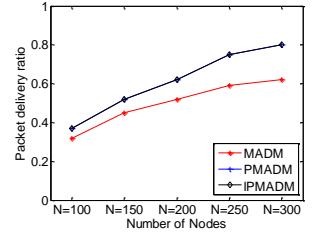

Fig. 9. Packet delivery ratio when the parameters vary slightly  
Fig. 10. Packet delivery ratio when the parameters vary greatly

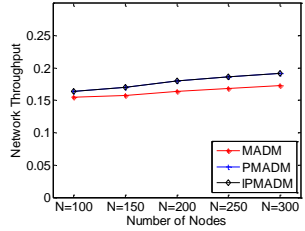 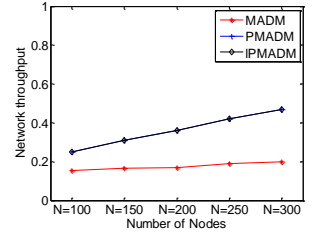

Fig. 11. Network throughput when the parameters vary slightly  
Fig. 12. Network throughput when the parameters vary greatly

The packet delivery ratio and network throughput are presented in Fig. 9, Fig. 10, Fig. 11 and Fig. 12, respectively. Similar to the transmission delay, with the increasing of the number of nodes, the packet delivery ratio and network throughput increase. On the one hand, the packet delivery ratio and the network throughput in PMADM and *l*PMADM algorithms are better than those in the MADM algorithm. On the other hand, when the parameters vary greatly, the performances of the PMADM and *l*PMADM algorithm are much better the MADM algorithm. For instance, for the network throughput, when the number of nodes is 250, the value in PMADM algorithm is 10% higher than that in the MADM algorithm in Fig. 11; however, this value is 54% in Fig. 12.

So, these simulation results demonstrate that the PMADM algorithm is effective, especially in the scenario that the parameter varies greatly. Note that the performances of the PMADM and *l*PMADM algorithm are the same. So, in these figures, the simulation results of the PMADM and lPMADM algorithm are coincident. The only difference between the PMADM algorithm and *l*PMADM algorithm is that the computational complexity of *l*PMADM algorithm is lower than the PMADM algorithm. This will be shown in Section VI.B.

### B. The computational complexity

In this section, we evaluate the computational complexity of these three algorithms. In this simulation, we use the times of utility computation to represent the computational complexity. The results are shown in Fig. 13.

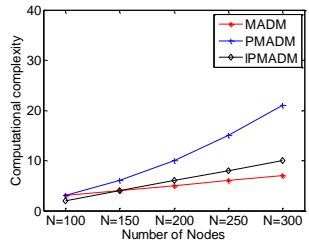

Fig. 13. The computational complexity of these three algorithms

In Fig. 13, we have two conclusions. First, with the increase of the number of nodes, the complexity of the proposed PMADM algorithm increases quickly; second, by using the $l$PMADM algorithm, the computational complexity of the PMADM algorithm is reduced greatly. Even the increasing in the $l$PMADM algorithm is still higher than that in the MADM algorithm, its increasing ratio is much smaller than that in the PMADM algorithm. Moreover, when the number of nodes in small, the complexity of $l$PMADM algorithm can be lower than that of the MADM algorithm.

## CONCLUSIONS

In this paper, we propose the pairwise-based MADM algorithm. In PMADM, only two nodes' utilities are calculated and compared at each time. The PMADM algorithm is much more accurate than the traditional MADM algorithm. Moreover, we also prove that the PMADM algorithm is sensitive to the parameter which varies seriously and in-sensitive to the parameter which changes slightly. This property is better than that of the traditional MADM algorithm. Moreover, the PMADM algorithm is stable than that of the traditional MADM algorithm. Moreover, for reducing the computational complexity of the PMADM algorithm, we propose the low-complexity PMADM algorithm. For analyzing the computational complexity of the $l$PMADM algorithm, we propose the tree-based decomposing algorithm in this paper. The $l$PMADM algorithm has the same properties and performances as the PMADM algorithm; however, it is simpler than that of PMADM algorithm. The theoretical analysis and the simulation results demonstrate that the proposed algorithms can improve the performances of MADM greatly.